# Comprehensive characterizing of vortex phases in type-II superconductor YBa$_2$Cu$_3$O$_{7-x}$ by a magnetoelectric technique

Peipei Lu[1,2,*], Jing Zhang[1,*], Jun Lu[3], Xiaoyuan Zhou[4], Young Sun[3,4,†] and Yisheng Chai[1,4,‡]

[1]Low Temperature Physics Laboratory, College of Physics, Chongqing University, Chongqing 401331, China

[2]College of Physics, Hebei Normal University, Shijiazhuang 050024, Hebei, China

[3] Beijing National Laboratory for Condensed Matter Physics, Institute of Physics, Chinese Academy of Sciences, Beijing 100190, China

[4]Center of Quantum Materials and Devices, Chongqing University, Chongqing 401331, China

[†]yschai@cqu.edu.cn

[‡]youngsun@cqu.edu.cn

[*] P. L. and J. Z. contributed equally to this work

**Abstract**

Vortex phases play a crucial role in determining the magnetic and electric properties of type-II superconductors. However, a universally applicable method for characterizing these vortex phases has been lacking. In this study, we present a comprehensive approach to studying vortex phases and phase boundaries in a $YBa_2Cu_3O_{7-x}$ polycrystalline sample, a type-II superconductor. Our method employs a magnetoelectric technique, wherein a thin piezoelectric material, $0.7Pb(Mg_{1/3}Nb_{2/3})O_3$-$0.3PbTiO_3$ (PMN-PT), is mechanically bonded to $YBa_2Cu_3O_{7-x}$, forming a laminate structure and acting as a strain gauge. Through this innovative approach, we successfully explore the phase diagram of the $YBa_2Cu_3O_{7-x}$ polycrystal. Notably, our technique accurately determines critical parameters such as $H_{c1}$, the irreversible line, $H_{c2}$, and distinguishes between vortex glass, vortex liquid, and non-vortex states. Furthermore, it enables dynamic response studies at varying frequencies, facilitating the observation of threshold phenomena within the vortex liquid phase. Additionally, our method enables the assessment of vortex density within the vortex solid phase. The applicability of our technique extends directly to the investigation of vortex phases in other type-II superconductors.

# I. INTRODUCTION

Superconductors (SC) exhibit two fundamental properties: zero resistivity and perfect diamagnetism, capable of fully excluding small external magnetic fields in principle [1-3]. The $\mu_0 H$ field inside the SC remains at zero until superconductivity is completely quenched, marked by the exceeding of a critical field, $H_{c1}$, in type-I SC. However, due to the negative interface energy between superconducting and normal phases in type-II SC, magnetic fields greater than $H_c$ can penetrate the SC phase instead of completely quenching it. Above $H_c$, magnetic fields manifest as quantized vortices in a type II SC, with each vortex carrying a quantum of $\phi_0 = hc/2e$. These vortices can arrange themselves into a regular triangular or square lattice, forming a two-dimensional solid state [4]. In an ideal two-dimensional crystal devoid of impurity/imperfection, the vortices will move under external current, losing zero resistivity [5]. In reality, impurities in the parent material can pin vortices, stabilizing their positions under small currents. Upon heating, introducing disorder, or applying high currents, the lattice phase can be melted into a liquid phase. This transition leads to finite resistance ($R \neq 0$) under a driven electric field [6], and it exhibits dissipative behavior (imaginary part of magnetic susceptibility $\chi'' \neq 0$) under a driven magnetic field [7]. In the liquid phase, vortices can detach from pinning centers beyond a certain magnitude of external stimulus, a phenomenon referred to as a threshold behavior [8]. The transition between vortex liquid and solid phases has been confirmed to occur at an irreversible temperature $T_{irr}$ [9]. Due to the quantum nature of vortices, they can serve as superconducting qubits in quantum computing [10] and potentially host

Majorana zero modes for topological quantum computing, as observed by STM/S in the vortex cores of FeTe$_{0.55}$Se$_{0.45}$ [11]. Furthermore, strong pinning in the vortex solid phase of type-II SC is of vital importance for its applications, particularly in areas such as superconducting magnets and planar magnetic field sources [12]. Hence, identifying vortex phases and comprehending their dynamic responses to external stimuli are crucial for their practical applications.

To identify vortex phases and study their dynamics, a comprehensive understanding requires the combination of various traditional methods. (1) Direct resistivity measurements only differentiate between the superconducting and normal phases, without resolving the vortex phases. Hence, large current density is necessary to measure the current-voltage relationship around Tc and to identify phase transitions between vortex glass and liquid, or normal and liquid phases [13]. (2) dc magnetization (*M*) can determine $T_{irr}$ through zero-field-cooling (ZFC) and field-cooling (FC) temperature scans [14]. However, for higher *H*, *M* becomes positive for both FC and ZFC, making it challenging to conclusively ascertain the presence of vortex phases. Furthermore, during magnetic field scans, the butterfly-shaped *M-H* loop with significant hysteresis is a typical character of the vortex solid phase due to strong pinning. (3) The ac susceptibility $\chi = \chi' + i\chi''$ serves as a sensitive and nondestructive approach for measuring SC properties [15]. The negative real part, $\chi'$, indicates superconducting diamagnetism. Meanwhile, the imaginary part, $\chi''$, implies the dissipative behavior of the vortex liquid phase due to depinning, while a $\chi''$ value of zero at lower temperatures points to a vortex solid phase. However, its real part only quantifies superconducting volume, lacking direct information about vortex density. (4) The magneto-optical [16] or Bitter technique [17] directly visualizes vortices or their slow dynamics. However, these methods require a high degree of sample surface smoothness. Additionally, they cannot directly determine vortex density in the bulk sample or capture rapid dynamics due to imaging speed limitations. (5) Neutron scattering experiments provide insights into the vortex lattice structure and its melting behaviors by directly revealing the reciprocal lattice through scattering patterns [18-20]. The diffraction signal vanishes when the sample enters the normal state. Nonetheless, this method needs a considerable amount of time for data collection and falls short in probing dynamic behaviors of the vortex liquid phase. (6) The magnetostriction method can also serve to detect properties of vortex phases [21]. Typically, vortex phases exhibit large magnetostriction owing to the pinning forces affecting vortices. Nonetheless, the slow measurement speed restricts the investigation of dynamic behaviors in the liquid phase. Hence, a method that is relatively simple, cost-effective, and rapid, while capable of providing a comprehensive characterization of vortex phases and phase boundaries, remains elusive.

Recently, a new method probing the magnetic skyrmion system, including the mixed and solid phases, is invented by using a composite magnetoelectric (ME) technique [22, 23]. In our experimental setup, a piezoelectric phase is mechanically bonded with a magnetostrictive phase to form a laminate structure, as schematically shown in Figure 1(a). The magnetostriction $\lambda = \Delta L/L$ of magnetic phase can be converted into an electrical signal via interfacial strain coupling, as depicted in Figure 1(b). Upon application of an ac magnetic field ($H_{ac3}$), an out-of-plane ac voltage ($V_{ac3}$) signal is generated in the piezoelectric phase with thickness t. The longitudinal ME voltage coefficient ($\alpha_{E33} = V_{ac3} / tH_{ac3}$) governing in-plane isotropic magnetostrictions in the piezoelectric material is described as follows [22-24]:

$$\alpha_{33} = \frac{V_{ac3}}{tH_{ac3}} \propto k d_{31} q_{13} \qquad (1)$$

where $k$ (0<$k$<1) is interface coupling parameter, $d_{ij}$ defines the transverse piezoelectric coefficient and $d_{31}=d_{32}$ for polycrystalline sample and $q_{13} = d\lambda_1/ dH_3$. In the case of MnSi single crystal samples, we successfully identify the presence of a skyrmion solid phase surrounded by mixed phases using this technique. Analogously, the two-dimensional particle system—the vortex phases in the type-II SC, usually have strong magnetostrictive behavior [21]. Consequently, this technique holds potential for the investigation of vortex phases in type-II superconductors.

In this study, we introduce an economical and simple composite ME technique to comprehensively investigate the vortex phases in the type-II high-temperature superconductor $YBa_2Cu_3O_{7-x}$ (YBCO) polycrystal. We demonstrate that this method enables a comprehensive exploration and plotting of the vortex phase diagram, including the $H_{c1}$, irreversible line, $H_{c2}$ and distinguish among the vortex glass, vortex liquid, non-vortex states. Furthermore, it can probe the dynamic response of the vortex liquid phase under various frequencies and stimulus strengths. Most importantly, it can quantitatively measure the vortex density of the vortex solid phase.

## II. EXPERIMENTAL

Commercial YBCO polycrystalline samples were acquired from the Central Iron & Steel Research Institute, China. These samples were fabricated using the melt-textured method, resulting in robust pinning forces exerted on the vortices. A standard four-probe method is applied in the resistance measurements. Prior to measurement, the YBCO polycrystal was annealed in air at 425 °C for 6 hours to reduce the contact resistance. Magnetization measurements were conducted using the VSM option of the Physical Property Measurement System (PPMS, Quantum Design, Dynacool). The ac susceptibility χ of YBCO was assessed using the ACMs option of the PPMS, employing various frequencies under an applied magnetic field strength $H_{ac}$ = 5 Oe.

The composite ME structure was assembled by bonding a YBCO polycrystalline sample to a [001]-cut single crystal of piezoelectric material $0.7Pb(Mg_{1/3}Nb_{2/3})O_3$–$0.3PbTiO_3$ (PMN-PT) ($t$ = 0.2 mm) using silver epoxy (Epo-Tek H20E, Epoxy Technology Inc.). Before conducting any electrical measurements, the PMN-PT undergoes pre-poling along the thickness direction through a poling electric field of 5.5 kV/cm for 1 hour at room temperature. An ac magnetic field (Hac) is generated using a coil, and the resulting ac ME voltage ($V_{ac}$ = $V_x$ + i$V_y$) across the YBCO/PMN-PT structure is detected using a lock-in amplifier (NF Corporation LI5645) with a commercial sample stick (MultiField Tech.), as depicted in Figure 1(c). The ac ME voltage coefficient (α = $V_{ac}/tH_{ac}$ = $α_x$ + i$α_y$) is subsequently computed. Temperature and dc magnetic field conditions are provided using the PPMS.

## III. RESULTS AND DISCUSSIONS

The superconductivity of the YBCO sample was initially characterized using conventional resistance and dc magnetization techniques. The resistance of YBCO as a function of temperature $R$(T) under different dc magnetic fields was measured, as shown in Figure 2(a). The characteristic temperatures $T_{on}$ and $T_{end}$ are defined as points where the resistance begins to decrease and becomes zero, respectively. Above $T_{on}$, YBCO is in its normal state, while below $T_{end}$, it enters the vortex glass phase due to the strong pinning and polycrystalline nature of the specimen. In the intermediate range, the material is anticipated to be in the vortex liquid phase. At zero magnetic field, the $T_{on}$ and $T_{end}$ transition temperatures are distinctly observed at approximately 91 K and 83 K, respectively. As the external magnetic field is increased, the transition temperatures are consistently suppressed and exhibit a slight broadening effect. Notably, two steps are evident between $T_{on}$ and $T_{end}$ in the $R$(T) curves, indicative of the existence of slush and liquid phases.

Temperature-dependent magnetization $M$(T) measurements were conducted during the field cooling (FC) and zero-field cooling (ZFC) processes under a 5 mT magnetic field, as depicted in Figure 2(b). Evident diamagnetism is observed below $T_M$ ≈ 89 K, and the superconducting volume fraction is estimated to be 48.4% at 60 K from the ZFC data. Furthermore, $T_M$ at this field corresponds to the irreversible line, marking the point where the FC and ZFC curves begin to merging. Figure 2(c) shows the $M$(T) curves under various dc magnetic fields during the FC process. The $T_M$ shifts to lower temperatures as the dc magnetic field increases. The $H$-$T$ phase diagram of the YBCO sample is presented in Figure 2(d). In this sample, $T_{on}$ and $T_M$ exhibit proximity, contrary to the typical scenario where the irreversible line and the $H_{c2}$ ($<T_{on}$) line are distinct in high-Tc cuprates. Therefore, from the above measurements and the derived phase diagram, it is very hard to pin down those lines separating the vortex solid, liquid, and normal states.

To confirm the effectiveness of our composite ME technique in investigating the vortex phases within the YBCO polycrystal sample, we constructed a composite ME laminate comprising a YBCO/PMN-PT plate configuration. Large ME voltage coefficients α were observed in the YBCO/PMN-PT laminate below $T_c$ when subjected to finite dc magnetic fields and a 1 Oe ac magnetic field with a frequency of 199 Hz, as depicted in Figure 3. The real part $α_x$ and imaginary part $α_y$ of α at 0 T (absent of vortices, not shown here) are featureless upon crossing $T_c$, thus they were regarded as background signals and subtracted from other data collected under finite magnetic fields. At 0.1 T for both ZFC and FC processes, $α_x$ values exhibit distinct separation at lower temperatures, and the FC data consistently surpasses the ZFC data, aligning with the behavior observed in *M* from Figure 2(b). At $T_{irr}$, two curves merge, coinciding with the definition of the so-called irreversible temperature. As temperature increases beyond this point, they exhibit a two-step decline and ultimately reach zero below $T_M$ and $T_{on}$. Conversely, at low temperatures below $T_{irr}$, $α_y$ values are negligible for both cases. Above $T_{irr}$, $α_y$ of both ZFC and FC exhibit closely matching temperature profiles characterized by two peaks centered at $T_{p1}$ = 84.1 K and $T_{p2}$ = 86.7 K below $T_M$, coinciding with the rapid, two-step decline observed in $α_x$. These dual step/peak characteristics further correspond to the two-step behavior observed in resistance, signifying the presence of two distinct transition temperatures within this sample. The temperature at which $α_y$ begins to approach zero is designated as $T_{ini}$, and this value is lower than both $T_M$ and $T_{on}$.

The presence of nonzero $α_x$ values below $T_M$ and $T_{on}$ strongly indicate the presence of vortex phases. In addition, the nonzero $α_y$ values observed between $T_{irr}$ and $T_{ini}$ indicate a dissipative behavior within the superconducting state, akin to the behavior observed in ac magnetic susceptibility [7]. The comparison between these two methods will be illustrated in Figure 4. The emergence of peaks in $α_y$ strongly indicates the presence of the vortex liquid phase within type-II superconductors, while the initial temperature $T_{ini}$ corresponds to the disappearance of the vortex phase. Below $T_{irr}$, the negligible $α_y$ values and the divergence between ZFC and FC signals in $α_x$ are in accord with the characteristics of a vortex glass phase. The smaller $α_x$ values observed during ZFC seem to imply a reduced vortex density, suggesting a positive correlation between the magnitude of $α_x$ and vortex density. This assertion is further supported by the observation that the $α_x$ signal vanishes when the vortex density is zero in the absence of a magnetic field.

In order to validate the aforementioned relationship, we measured $α_x$ and $α_y$ under various dc magnetic fields during the FC process: 0.01, 0.02, 0.05, 0.1, 0.2, 0.5, 1, 2, 5, and 9 T. The corresponding results are presented in Figure 3(c) and 3(d). With increasing the magnetic field, the magnitude of $α_x$ increases. Moreover, the $T_{irr}$ is suppressed continuously, indicating the bending of irreversible line under high field. In the curves of $α_y$, a negligible ME signal is observed below the $T_{irr}$ temperature for each dc magnetic field, affirming the presence of the vortex glass phase beneath this temperature. At 60 K, the values of $\alpha_x$ exhibit an almost linear proportionality with the dc magnetic field, as depicted in the inset of Figure 3(c). Typically, the dc magnetic field is proportional to the vortex density within the vortex solid phase during the FC process, confirming the linear correlation between $\alpha_x$ and the vortex density. The relationship is different from magnetization and ac magnetic susceptibility measurements wherein lower vortex density results in more pronounced diamagnetic signals [7, 14].

Within the curves of $\alpha_y$, $T_{p1}$, $T_{p2}$, and $T_{ini}$ are progressively suppressed to lower temperatures, accompanied by broader intervals at higher magnetic fields. This phenomenon suggests that the vortex liquid phase has been shifted to lower temperatures. To prove the association between the non-zero features in $\alpha_y$ and the vortex liquid phase, we directly compared the ac susceptibility and the ac ME voltage coefficient α at 2 T at selected frequencies, as shown in Figure 4. Concerning the real part, both the ac susceptibility χ' (Figure 4(a)) and $\alpha_x$ (Figure 4(c)) converge to zero at the same temperature, $T_{irr}$, although the distinct two-step characteristic is more pronounced in $\alpha_x$ than in χ'. In the imaginary part, both χ'' (Figure 4(b)) and $\alpha_y$ (Figure 4(d)) converge to zero at the same temperature, $T_{ini}$, exhibiting three-peak and two-peak features, respectively. The two higher peaks manifest almost identical temperatures across both approaches, while the lower $T_{p3}$ is solely discernible in the ac susceptibility χ''. Furthermore, all three peaks shift towards higher temperatures as the frequency is elevated, as demonstrated in Figure 4(b) and 4(d). To extract the activation energy ($E_a$) and characteristic relaxation time ($\tau_0$), each peak is fitted using the Arrhenius equation, as depicted in Figure 4(e) [25].

$$\ln f = \ln \frac{1}{\tau_0} - \frac{E_a}{RT} \qquad (2)$$

where R is gas constant. The calculated activation energies (Ea) for $T_{p1}$ and $T_{p2}$, obtained from the slopes using both the ME voltage technique (20.8 and 25 meV) and ac susceptibility (22.9 and 32.3 meV), exhibit a remarkable concurrence. The emergence of a third peak in χ'' could potentially stem from the larger excitation field of 5 Oe. This striking resemblance observed between the imaginary components of the two methods corroborates the presence of the vortex liquid phase existing between $T_{irr}$ and $T_{ini}$.

To gain deeper insights into the relationship between vortex density and the real part of the ME signal $\alpha_x$, *M* and α as a function of magnetic field after ZFC process at 10 and 5 K were measured, respectively, as shown in Figure 5. A quasi-linear diamagnetic feature is evident in the *M(H)* curve below the critical field $H_{c1}$. Beyond $H_{c1}$ up to 9 T, the magnitude of *M* diminishes with increasing *H*, indicative of the fast enhancement of vortex density within the vortex glass phase. As the magnetic field starts to down sweep from 9 T, a swift reversal of sign in *M* is induced, seemingly suggesting a decline in vortex density. Subsequently, within the range from 8 T to -9 T, *M* peaks around zero *H*, signifying a transition of vortex density from nonzero to zero and subsequently back to a high density state. The *M(H)* loop as a whole exhibits pronounced hysteresis, attributed to the demagnetization effect. In contrast, the real part of the ME signal, $\alpha_x$, demonstrates a near-linear increment or decrement with respect to H and displays minimal hysteresis behavior, while the imaginary part, $\alpha_y$, remains negligible within the studied field range, as illustrated in Figure 5(b). In the low-field region following the ZFC process, the $\alpha_x$ slightly bends up with a larger slope after $H_{c1}$, as shown in Figure 5(c). This is consistent with the notion of an accelerated increase in vortex density subsequent to this critical field. Nonetheless, our data unveil distinct vortex dynamics within the vortex glass phase. 1) The linear and nonzero behavior of $\alpha_x$ below $H_{c1}$ suggests the continual entrance of magnetic vortex lines into the sample, indicative of an imperfect Meissner state. 2) When the magnetic field is initially reduced from 9 T, $\alpha_x$ (representing the vortex density) experiences a slight increase, contrary to the significant decrease implied by the $M(H)$ curve. 3) During both upward and downward magnetic field sweeps, the behavior of the vortex density appears to closely reflect the trend without substantial hysteresis, evident from the feeble hysteresis observed in the $\alpha_x$ $(H)$ curve. Hence, the real part of the ME signal, $\alpha_x$, appears capable of providing a more direct measure to the vortex density within the type-II superconductor YBCO.

Finally, we compile all the temperature scan data to construct a comprehensive phase diagram for the YBCO polycrystal, presented in Figure 6(a). Through a comparative analysis of characteristic temperatures across different techniques, four distinct phases are identified: vortex glass, vortex liquid, vortex cluster, and normal states. The temperatures $T_{irr}$ and $T_{end}$ define the irreversible line separating the vortex glass and vortex liquid phases. $T_{ini}$ designates the transition temperature between the vortex liquid phase and cluster state, wherein the superconducting liquid coexists with the normal state without vortices. Meanwhile, $T_M$ and $T_{on}$ determine the transition temperature between the cluster and normal states, deviating from the conventional irreversible line or $H_{c2}$ line, possibly attributed to sample inhomogeneity. To corroborate this proposed phase diagram, we systematically varied the magnitude of the applied $H_{ac}$ and recorded the corresponding ME voltage, $V_{ac} = V_x + iV_y$, at temperatures of 70, 77.5, 81, and 90 K under a magnetic field of 2 T, as illustrated in Figure 6(b) and 6(c). In the normal state (at 90 K), both $V_x$ and $V_y$ remain nearly zero across different $H_{ac}$ values. Conversely, within the vortex liquid phase (at 81 K), $V_y$ deviates from its zero value, exhibiting rapid growth, while $V_x$ experiences a swift departure from linear behavior under very slight $H_{ac} \approx 0.1$ Oe. This deviation signifies a subtle threshold response characteristic of the liquid nature. Within the vicinity of the vortex glass phase (at 77.5 K), the threshold value of $H_{ac}$ approximates 1 Oe, indicative of stronger pinning effects from impurities. In the deep vortex glass phase (at 70 K), the threshold value of $H_{ac}$ surpasses 2 Oe, accompanied by $V_x$ maintaining a strictly linear relationship with $H$. In light of these findings, our ME technique reliably establishes the vortex-related phase diagram with enhanced clarity compared to conventional methods.

Need to point out that the positive relationship between $\alpha_x$ ($\propto \Delta L/L\Delta H$) and the density of vortex is counterintuitive. Usually, the number of vortices in a SC should be proportional to the changes in sample geometry that the $\Delta L/L$ should have a linear relationship with the density of vortex. However, the observed behavior of $\alpha_x$ is far from constant, contrary to the above expectations. One plausible explanation for this discrepancy is that the application of an ac driven field can induce low-frequency eddy currents, setting the vortex ensemble in motion within the sample and thereby causing a torsional deformation of the ME composite. This deformation is likely proportional to the vortex density.

Last but not least, we would like to outline several key technical advantages of our proposed method. (1) The accuracy achieved is high and noise levels are kept low, owing to the utilization of the lock-in technique. The lock-in amplifier sets the smallest detectable change in ME voltage at approximately ±0.1 μV, as demonstrated in Figure 3(b). (2) The cost of the measurement remains economical, largely attributed to the affordability and accessibility of the PMN-PT plate. Furthermore, the measurement setup itself is cost-effective, requiring minimal additional components—specifically, an extra lock-in amplifier and a current source. (3) The sample preparation process is fast and simple, necessitating only the bonding of the sample with the PMN-PT plate, along with a short pre-poling time of approximately 2 hours. (4) Notably, the measurement speed is rapid, operating at a rate of 1 second per data point—significantly quicker than the ac susceptibility method, which often involves intricate compensation processes. (5) Moreover, the test frequency range surpasses that of ac susceptibility, spanning from 1 Hz to 100 kHz, enhancing the versatility and scope of our proposed technique.

## IV.   CONCLUSION

This study successfully introduces a novel ME technique utilizing the YBCO/PMN-PT structure for characterizing the vortex phases in the polycrystalline type-II superconductor YBCO. From the characteristic temperatures found in ME signals, the distinctive vortex glass, vortex liquid, and non-vortex states are discerned, corresponding to the definition of $H_{c1}$, irreversible, and $H_{c2}$ lines. Furthermore, the technique enables the investigation of dynamic responses within the vortex liquid phase under various frequencies and stimulus strengths. Most important, the study unveils a quantitative correlation between the real part of the ME signal $α_x$ and the vortex density. Overall, this composite magnetoelectric technique opens a promising avenue for exploring vortex dynamics across various type-II superconductors.

## ACKNOWLEDGEMENTS

This work was supported by the National Natural Science Foundation of China (Grant Nos. 11974065, 52101221), and the National Key Research and Development Program of China (Grant No. 2016YFA0300700), Chongqing Research Program of Basic Research and Frontier Technology, China (Grant No. cstc2020jcyj-msxmX0263), Fundamental Research Funds for the Central Universities, China (2020CDJQY-A056, 2020CDJ-LHZZ-010, 2020CDJQY-Z006), the Natural Science Foundation of Hebei Province (Grant No. A2021205022), Science and Technology Project of Hebei Education Department (QN2021088), Hebei Normal University (Grant No. L2021B09). Y. S. Chai would like to thank the support from Beijing National Laboratory for Condensed Matter Physics. We would like to thank Miss G. W. Wang at Analytical and Testing Center of Chongqing University for her assistance.

**References**

1. H. K. Onnes, Commun. Phys. Lab. Univ., Leiden, 122 (1911).

2. H. K. Onnes and W. Tuyn, Commun. Leiden Suppl., 50a (1924).

3. W. Meissner and R. Ochsenfeld, Naturwissenschaften, **21**, 787 (1933).

4. G. Blatter, M. V. Feigel'Man, and V. B. Geshkenbein, Rev. Mod. Phys., **66,** 1125 (1994).

5. H. K. Onnes, Commun, Phys. Lab. Univ. Leiden, Suppl.,34 (1913).

6. J. A. Fendrich, W. K. Kwok, J. Giapintzakis, C. J. van der Beck, V. M. Vinokur, S. Fleshier, U. Welp, H. K. Viswanathan, and G. W. Crabtree, Phys. Rev. Lett., **74**, 7 (1995).

7. A. Crisan, Y. Tanaka, D. D. Shivagan, A. Iyo, L. Cosereanu, K. Tokiwa, and T. Watanabe, Jpn. J. Appl. Phys., **46**, L451 (2007).

8. C. Reichhardt and C. J. Olson Reichhardt, Rep. Prog. Phys., **80**, 026501 (2017).

9. P. Chaddah, Pramana-J. Phys., **36**, 353 (1991).

10. M. Kjaergaard, M. E. Schwartz, J. Braumüller, P. Krantz, Joel I.-J.Wang, S. Gustavsson, and W. D. Oliver, Annu. Rev. Condens. Matter Phys., **11,** 369 (2020).

11. S. D. Sarma, M. Freedman and C, Nayak, npj Quantum Information, **1,** 15001 (2015).

12. J. H. Lacy, A. Cridland, J. Pinder, A. Uribe, R. Willetts, and J. Verdu, IEEE Transactions on Applied Superconductivity, **30**, 8 (2020).

13. Y. Liu, Y. S. Chai, H. J Kim, G. R. Stewart, and K. H. Kim, J. Korean Phys. Soc, **55**, 383 (2009).

14. A. Galluzzi, A. Nigro, R. Fittipaldi, A. Guarino, S. Pace, and M. Polichetti, J. Magn. Magn. Mater., **475**, 125 (2019).

15. K. H. Muller, Physica C, **159**, 717 (1989).

16. P. E. Goa, H. Hauglin, M. Baziljevich, E. Il'yashenko, P. L. Gammel and T. H. Johansen, Supercond. Sci. Technol., **14**, 729 (2001).

17. S. Ohshima, T. Kawai, S. Kanno, and H. Yamada, Physica C, **372-376**, 1872 (2002).

18. N. Rosov, J. W. Lynn, and T. E. Grigereita, J. Appl. Phys., **76**, 15 (1994).

19. M. R. Eskildsen, L. Ya. Vinnikov, T. D. Blasius, I. S. Veshchunov, T. M. Artemova, J. M. Densmore, C. D. Dewhurst, N. Ni, A. Kreyssig, S. L. Bud'ko, P. C. Canfield, and A. I. Goldman, Phys. Rev. B, **79**, 100501R (2009).

20. J. M. Densmore, P. Das, K. Rovira, T. D. Blasius, L. DeBeer-Schmitt, N. Jenkins, D. McK. Paul, C. D. Dewhurst, S. L. Bud'ko, P. C. Canfield, and M. R. Eskildsen, Phys. Rev. B, **79**, 174522 (2009).

21. A. Nabialek, H. Szymczak, and V. V. Chabanenko, J. Low Temp. Phys., **139**, 309 (2005).

22. M. I. Bichurin and V. M. Petrov, Phys. Rev. B, **68**, 054402 (2003).

23. G. Srinivasan, Annu. Rev. Mater. Res., **40**, 153 (2010).

24. D. Patil, June-Hee Kim, Y. S. Chai, Joong-Hee Nam, J. H. Cho, B. I. Kim and K. H. Kim, Appl. Phys. Express, **4**, 073001 (2011).

25. H. Salamati and P. Kameli, J. Magn. Magn. Mater., **278**, 237 (2004).

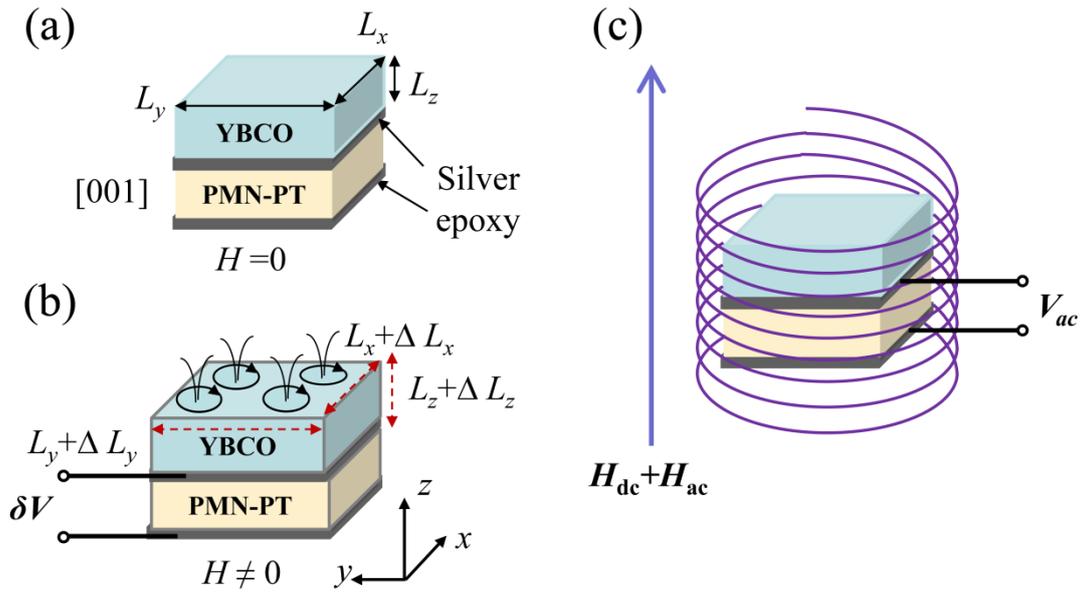

**Figure 1.** (a) The composite structure of the type-II superconductor polycrystalline YBCO and PMN-PT under zero magnetic field. (b) The shape changes of the YBCO sample in the composite structure under a magnetic field. (c) The magnetoelectric voltage $V_{ac}$ measured under an ac magnetic field by a magnetoelectric technique.

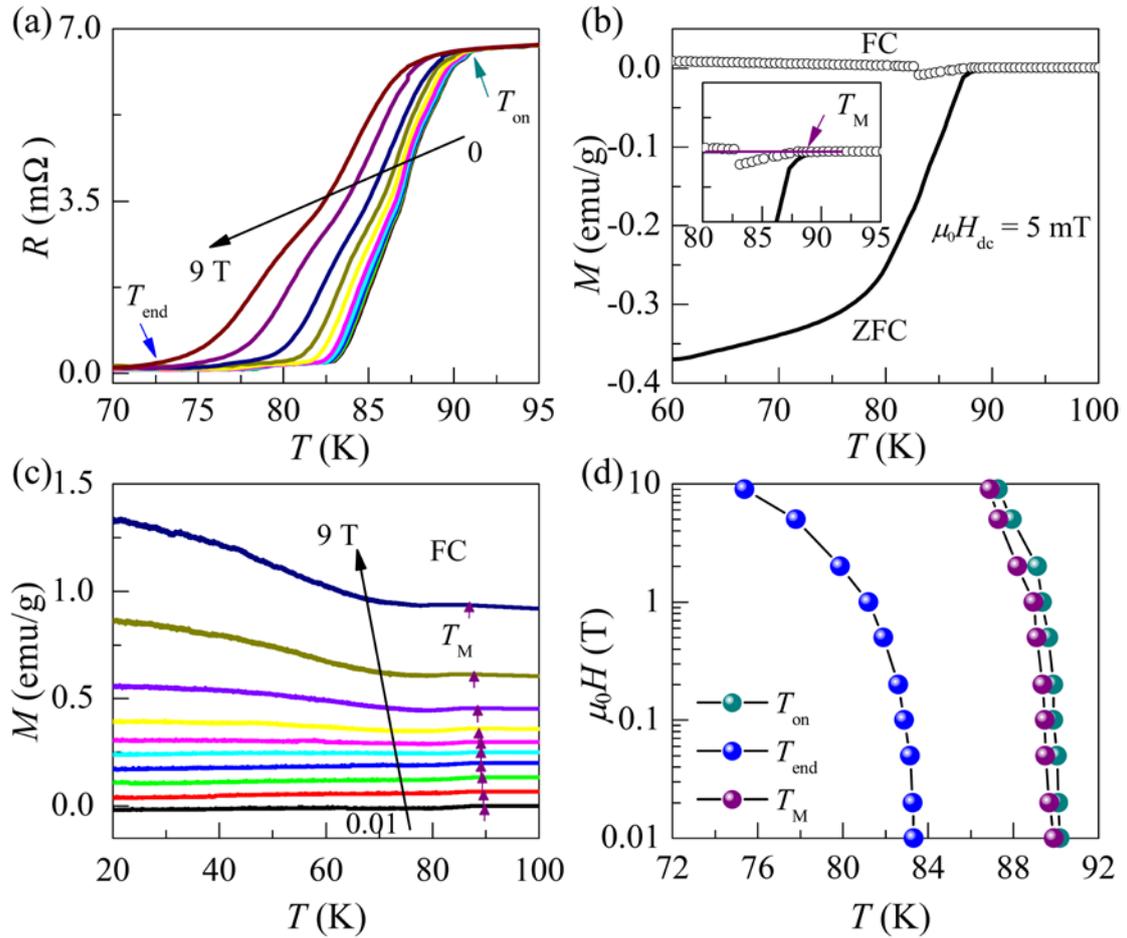

**Figure 2.** (a) Temperature dependent resistance curves under various magnetic fields of 0, 0.01, 0.02, 0.05, 0.1, 0.2, 0.5, 1, 2, 5 and 9 T for YBCO polycrystalline. (b) Temperature dependence of magnetization under a magnetic field of 5 mT. Both field-cooling (FC) and zero-field-cooling (ZFC) are shown. (c) Temperature dependent magnetization measured in FC condition at 0.01, 0.02, 0.05, 0.1, 0.2, 0.5, 1, 2, 5 and 9 T. (d) The phase diagram of the YBCO polycrystalline according to the data of (a) and (c).

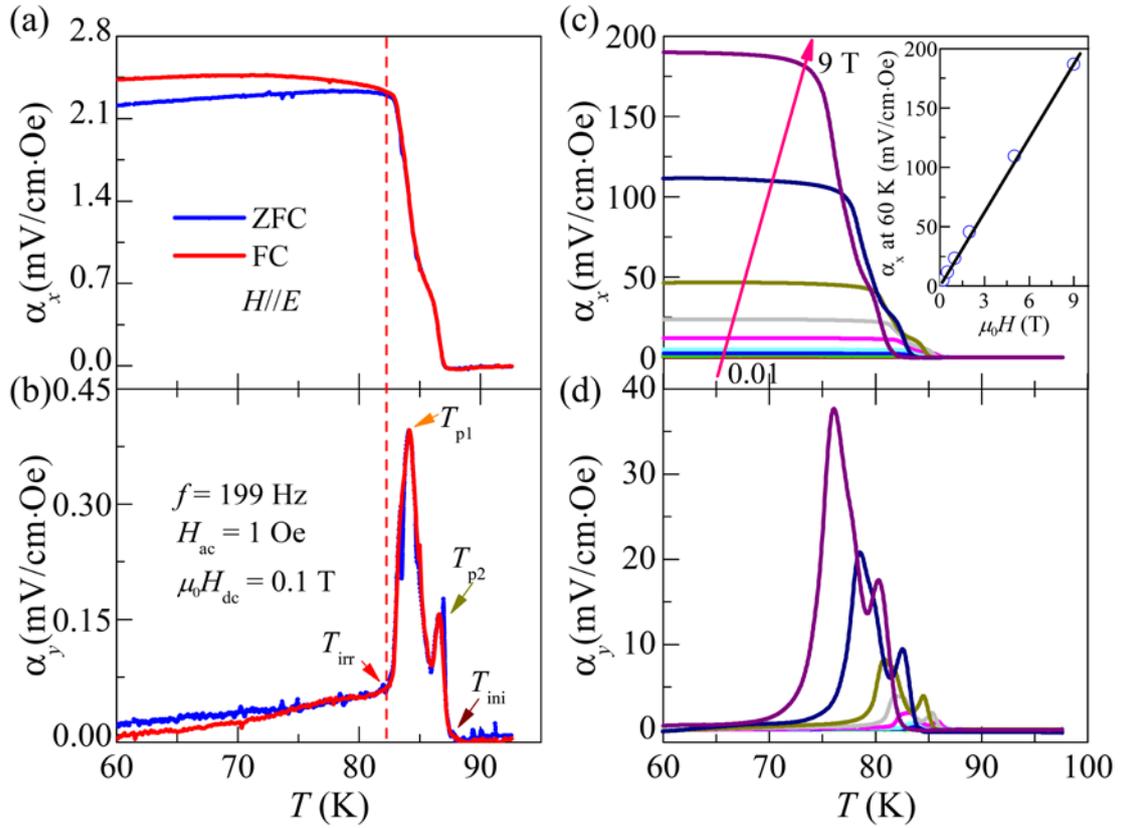

**Figure 3.** Under an out of plane ac magnetic field of 1 Oe, with a frequency of 199 Hz, the ac ME coefficient $\alpha_x$ (a) and $\alpha_y$ (b) for ZFC and FC processes at 0.1 T. $\alpha_x$ (c) and $\alpha_y$ (d) as a function of temperature for FC process of the YBCO/PMN-PT structure under different magnetic fields of 0.01, 0.02, 0.05, 0.1, 0.2, 0.5, 1, 2, 5 and 9 T.

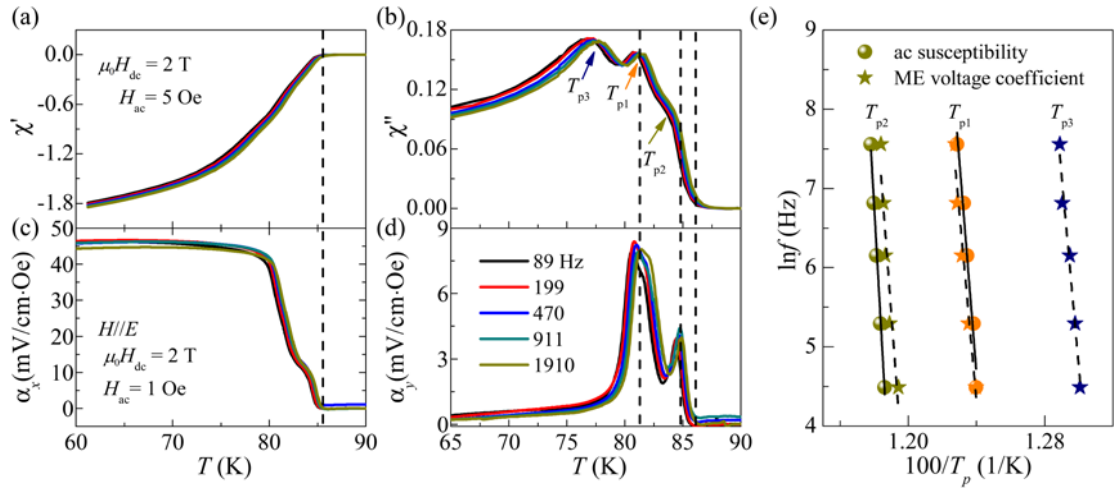

**Figure 4.** With a dc magnetic field of 2 T and selected frequencies of 89, 199, 470, 997 and 1910 Hz, the ac susceptibility χ' (a) and χ'' (b) under the ac magnetic field of 5 Oe for YBCO polycrystal, the ac ME coefficient $\alpha_x$ (c) and $\alpha_y$ (d) under an ac magnetic field of 1 Oe for YBCO/PMN-PT structure as a function of temperature. (e) The three peak temperatures as a function of frequency in Arrhenius plot.

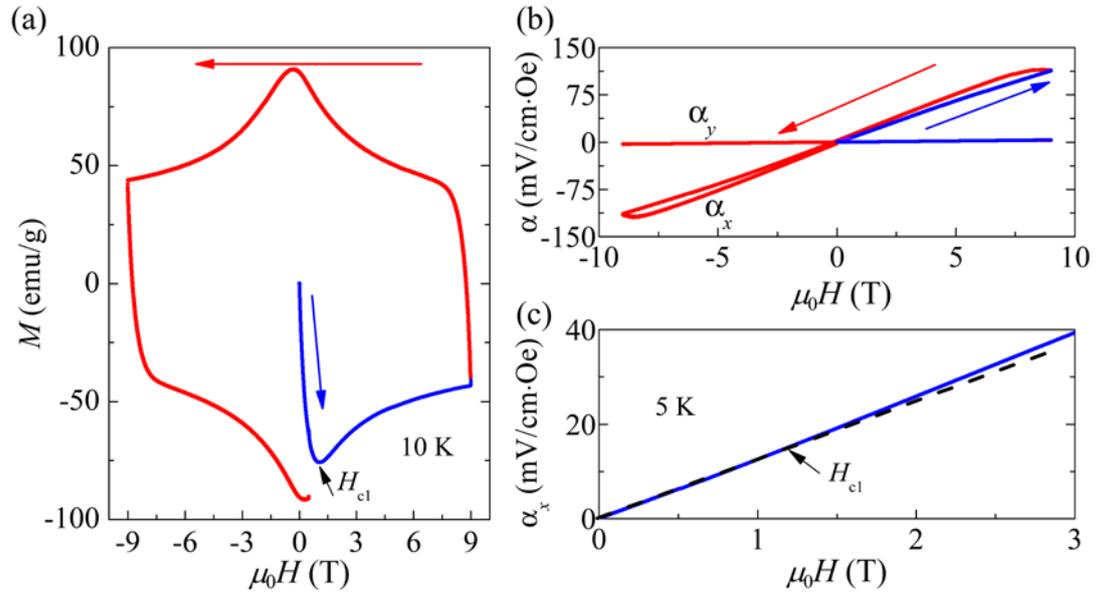

**Figure 5.** (a) The *M-H* loop of YBCO polycrystalline. (b) The ac ME coefficient $\alpha_x$ and $\alpha_y$ (b) as a function of magnetic field for YBCO/PMN-PT composite at 5 K. (c) The enlarged view of $\alpha_x$ in (b). $\alpha_x$ deviates from the linear after $H_{c1}$.

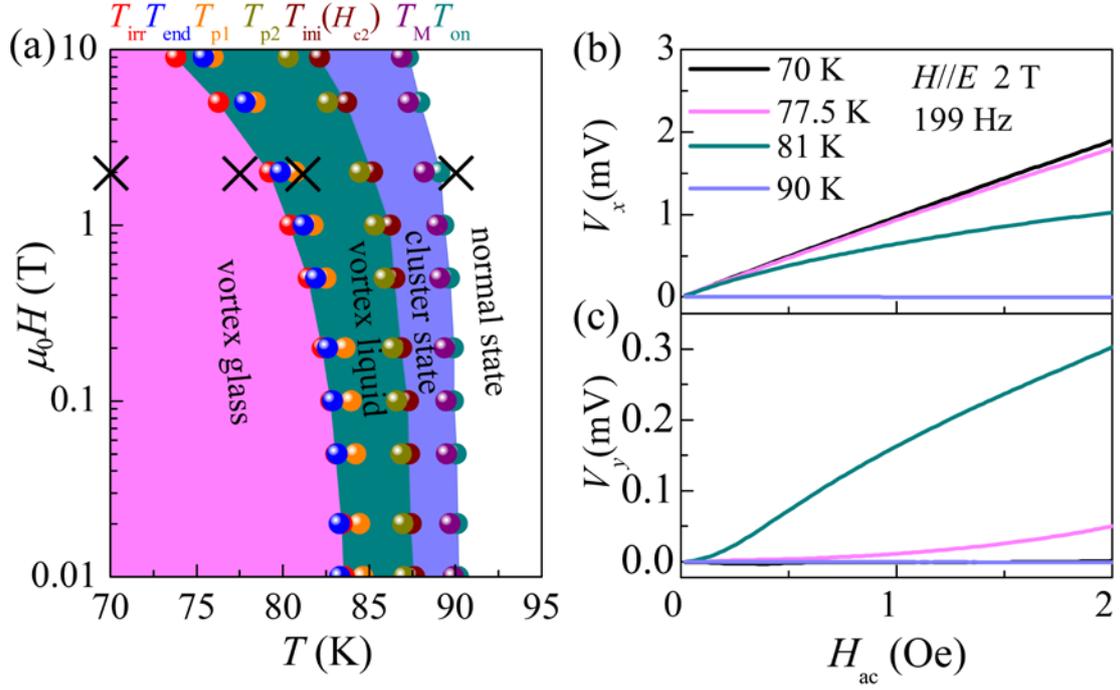

**Figure 6.** (a) The phase diagram of polycrystalline YBCO from the temperature dependent ME signals of YBCO/PMN-PT structure by magnetoelectric technique. Under a dc magnetic field of 2 T, the ac ME voltage $V_x$ (b) and $V_y$ (c) as a function of $H_{ac}$ at selected temperatures of 70, 77.5, 81 and 90 K.